\documentclass[preprint,1p,12pt]{elsarticle}

\usepackage{graphics}
\usepackage{graphicx}
\usepackage{subfigure}
\usepackage{textcomp}
\usepackage{amssymb}

\usepackage{vmargin}
\usepackage{float}
\usepackage{psfrag}
\usepackage[toc,page]{appendix} 
\usepackage{enumitem}

\journal{Journal of Sound and Vibration}

\begin{document}

\begin{frontmatter}



\title{Enhancing the Dynamic Range of Targeted Energy Transfer in Acoustics Using Several Nonlinear Membrane Absorbers}


\author[lma]{R. Bellet}
\cortext[cor]{corresponding author.}
\ead{bellet@lma.cnrs-mrs.fr}
\author[ecm]{B. Cochelin\corref{cor}}
\ead{bruno.cochelin@ec-marseille.fr}
\author[amu]{R. C\^ote}
\ead{cote@lma.cnrs-mrs.fr}
\author[lma]{P.-O. Mattei}
\ead{mattei@lma.cnrs-mrs.fr}

\address[lma]{CNRS-LMA, UPR 7051, F-13402 Marseille Cedex 20, France}
\address[ecm]{\'Ecole Centrale Marseille, CNRS-LMA UPR 7051, F-13451 Marseille Cedex, France}
\address[amu]{Aix-Marseille Univ, CNRS-LMA UPR 7051, F-13402 Marseille Cedex 20, France}
\begin{abstract}

In order to enhance the robustness and the energy range of efficiency of targeted energy transfer (TET) phenomena in acoustics, we discuss in this paper about the use of multiple nonlinear membrane absorbers in parallel. We show this way, mainly thanks to an experimental set-up with two membranes, that the different absorbers have additional effects that extend the efficiency and the possibilities of observation of TET. More precisely, we present the different behavior of the system under sinusoidal forcing and free oscillations, characterizing the phenomena for all input energies. The frequency responses are also presented, showing successive clipping of the original resonance peak of the system. A model is finally used to generalize these results to more than two NES and to simulate the case of several very similar membranes in parallel which shows how to extend the existence zone of TET.

\end{abstract}


\end{frontmatter}

\section{Introduction}

It has been shown that the use of nonlinear absorber (or nonlinear energy sink) in the field of passive vibration control can provide very interesting results, exploiting the targeted energy transfer (TET) phenomenon. The principle is to couple together a primary system and a nonlinear energy sink (NES) in order to achieve an irreversible transfer of energy towards the absorber. The dynamics of such a system differs dramatically from those of linear absorbers and have been described in details in \cite{gendelman01,vakakis01,vakakis01_2,vakakis03,vakakis04} in terms of resonance capture and nonlinear normal modes. A lot of papers have been published about different cases of linear system (a wave guide \cite{vakakis05}, a rod \cite{georgiades07,panagopoulos04,panagopoulos07}, a beam \cite{georgiades07_2}, a plate \cite{georgiades09}, a chain of coupled linear oscillators \cite{manevitch03}) and different applications (seismic mitigation \cite{gourdon07,nucera08}, aeroelastic instabilities control \cite{lee07,lee07_2} or drill-string systems stabilization \cite{viguie09}). For applications to Acoustics, the possibility of using a NES (composed by a thin circular viscoelastic membrane) coupled to an acoustic medium was demonstrated in \cite{cochelin06,bellet10} as a new technique of passive noise control in the low frequency domain where no efficient dissipative mechanism exists.

In every case, one of the drawbacks of targeted energy transfer mechanism is the existence of limited energy range in which the NES has its maximal efficiency. In order to enhance this range, it has been demonstrated in \cite{gourdon05,lee08} that the use of several NES could be a good solution. As in most papers about TET, the considered system is mechanical and the configuration is the nongrounded one, where the NES is only attached to the primary system, via a nonlinear stiffness. The authors showed that the most efficient configuration for the multiple NES is a series coupling, as a parallel coupling creates problems of additional mass effect. Indeed, in a nongrounded configuration, the NES has to be as light as possible. In the case of controlling an acoustic medium, the only possible configuration is grounded (the membrane has to be attached to a wall), the additional mass effect is not a problem and a parallel NES configuration is easier to build than a series one. We therefore chose to work on the use of several nonlinear membrane absorbers in parallel to control an acoustic primary system in order to enhance the robustness and the dynamic range of TET phenomena.

In this study, we begin with measurements of the response of a 2 NES system. The NES have very different activation thresholds, which permits their individual monitoring. Then we simulate similar systems, with 2 or 4 membranes. Then we simulate a different situation where NES properties are very close, and we analyze our results. The paper is organized as follows : first we present the setup and the calculation model. Then we present and analyze measurements made with 3 different excitation regimes : sine source, free oscillations, and swept sine source. The third part is devoted to numerical simulations, and then we conclude.

\section {Experimental study}
\subsection {Experimental set-up}

\begin{figure}
\centering
\includegraphics[width=6cm]{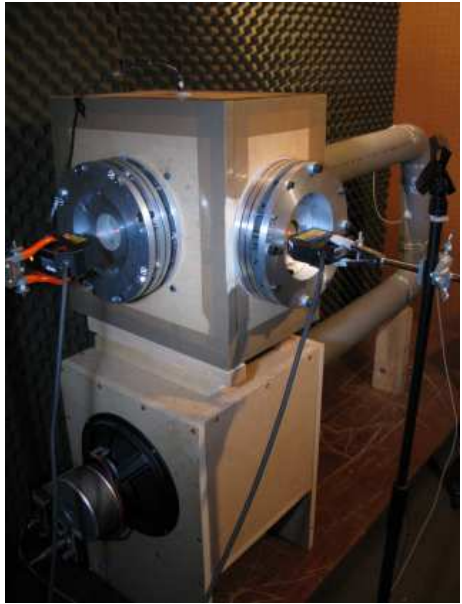}
\caption{Experimental set-up with two membrane absorbers.}
\label{pic}
\end{figure}

The experimental set-up is based on the set-up presented in \cite{bellet10}, with an acoustic source made of a loudspeaker and a box, a primary acoutic system based on the first mode of an open tube and a nonlinear absorber made of a thin circular visco-elastic membrane, coupled to the tube by a coupling box. As the acoustic pressure is spatially uniform in the coupling box, the side on which the membrane is settled does not matter. We thus used one of the free faces of this box to add a second membrane, leading to a configuration with two nonlinear membrane absorbers in parallel. Except for this second membrane, the set-up is identical to the one fully described in reference \cite{bellet10}. Two displacement sensors (laser triangulation) are used to measure the displacement of the center of the membranes (see Fig. \ref{pic}). The displacements are positives for inward movements of the membranes.
\begin{figure}
\centering
\includegraphics[width=14cm]{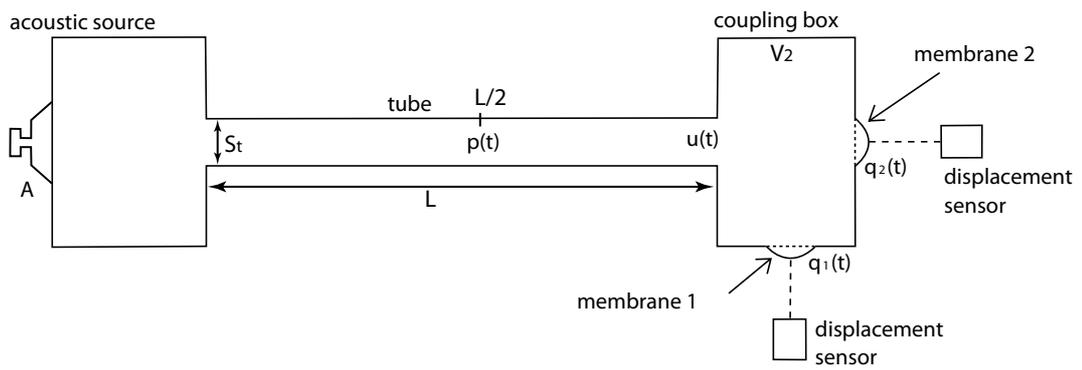}
\caption{Scheme of the experimental set-up with two membrane absorbers. $\cal A$: amplitude voltage of the sinusoidal input signal. $p(t)$: acoustic pressure at the middle of the tube. $u(t)$: air displacement at the end of the tube. $q_1(t)$ and $q_2(t)$: respective displacements of the centers of the membranes 1 and 2.}
\label{scheme}
\end{figure}

The model associated to the set-up is also the same than the previous one, with an additional membrane equation, leading to a three degrees of freedom (3dof) system:
\begin{eqnarray}\label{eq:systdim2m}
&&m_a \ddot{u} + c_f \dot{u} + k_a u + S_t k_b (S_t u - \frac{S_1}{2} q_1 - \frac{S_2}{2} q_2) = F cos(\Omega t) \nonumber\\ 
&&m_1 \ddot{q}_1 + k_{11} [ (\frac{f_{11}}{f_{01}})^2 q_1 + \eta_1 \dot{q}_1 ] + k_{31} [q^3_1 + 2 \eta_1 q^2_1 \dot{q}_1 ] + \frac{S_1}{2} k_{b} (\frac{S_1}{2} q_1 + \frac{S_2}{2} q_2 - S_t u) = 0 \nonumber\\
&&m_2 \ddot{q}_2 + k_{12} [ (\frac{f_{12}}{f_{02}})^2 q_2 + \eta_2 \dot{q}_2 ] + k_{32} [q^3_2 + 2 \eta_2 q^2_2 \dot{q}_2 ] + \frac{S_2}{2} k_{b} (\frac{S_1}{2} q_1 + \frac{S_2}{2} q_2 - S_t u) = 0 \nonumber\\
&&\ \ \ \ \ \ \ \ \ \mbox{with}\ \ \ m_a=\frac{\rho_a S_t L}{2} \ ,\ \ m_i= \frac{\rho_m h S_i }{3} \ ,\ \ k_b=\frac{\rho_a c_0^2}{V_2} \ ,\ \ k_a=\frac{\rho_a S_t c_0^2 \pi^2}{2L} \nonumber\\
&&\ \ \ \ \ \ \ \ \ \ \ \ \ \ \ k_{1i}=\frac{1.015^4 \pi^5}{36}\frac{Eh_i^3}{(1-\nu^2)R_i^2} \ ,\ \ k_{3i}=\frac{8\pi E h_i}{3(1-\nu^2) R_i^2} \nonumber\\
&&\ \ \ \ \ \ \ \ \ \ \ \ \ \ \ f_{0i}=\frac{1}{2\pi} \sqrt{\frac{1.015^4 \pi^4}{12}\frac{Eh_i^2}{(1-\nu^2)\rho_m R_i^4}} \ ,\ \ S_i=\pi R_i^2 \nonumber
\end{eqnarray}
In that system, the 3dof are the three displacements $u(t), q_1(t), q_2(t)$ of the air at the end of the tube and the center of the membranes $M_1$ and $M_2$ respectively. The other parameters are:

\begin{tabular}{ll}
   $\rho_a=1.3$ kg.m$^{-3}$ & density of the air \\
   $S_t=6.9 \times 10^{-3}$ m$^2$ & section of the tube\\
   $L=2$ m & length of the tube\\
   $\rho_m=980$ kg.m$^{-3}$ & density of the membranes\\
   $h_1=0.18$ mm & thickness of $M_1$\\
   $h_2=0.39$ mm & thickness of $M_2$\\
   $R_1=20$ mm & radius of  $M_1$\\
   $R_2=30$ mm & radius of $M_2$\\
   $S_1=1.3 \times 10^{-3}$ m$^2$ & surface of $M_1$\\
   $S_2=2.8 \times 10^{-3}$ m$^2$ & surface of $M_2$\\
   $f_{11}=59$ Hz & first resonance frequency of $M_1$ with a certain pre-stress\\
   $f_{12}=47$ Hz & first resonance frequency of $M_2$ with a certain pre-stress\\
   $f_{0i}$ & first resonance frequency of $M_i$ without pre-stress\\
   $E=1.48$ MPa & Young modulus of the membranes\\
   $\nu=0.49$ & Poisson's ratio of the membranes\\
   $\eta_i$ & damping factor of $M_i$\\
   $F$ & amplitude of the excitation term\\
   $V_2=27 \times 10^{-3}$ m$^3$ & volume of the coupling box\\
\end{tabular}
\\ \ 
\\ \
The different energies are then given by:
\begin{eqnarray}\label{eq:E}
&& E_{tube} = \frac{1}{2}m_a \dot{u}^2 + \frac{1}{2}k_a u^2 \nonumber\\ 
&& E_{membranes} = \frac{1}{2}m_1 \dot{q}^2_1 + \frac{1}{2}k_{11}(\frac{f_{11}}{f_{01}})^2 q^2_1 + \frac{1}{4}k_{31} q^2_1 + \frac{1}{2}m_2 \dot{q}^2_2 + \frac{1}{2}k_{12}(\frac{f_{12}}{f_{02}})^2 q^2_2 + \frac{1}{4}k_{32} q^2_2 \nonumber\\ 
&& E_{coupling\ box} = \frac{1}{2}k_b (S_t u - \frac{S_1}{2} q_1 - \frac{S_2}{2} q_2)^2 \nonumber\\ 
&& E_{total} = E_{tube} + E_{membranes} + E_{coupling box} \nonumber
\end{eqnarray}

\subsection {Regimes under sinusoidal excitation}

\begin{figure}
\centering
\subfigure[${\cal A}=0.05\ V$.]{\includegraphics[width=6.5cm]{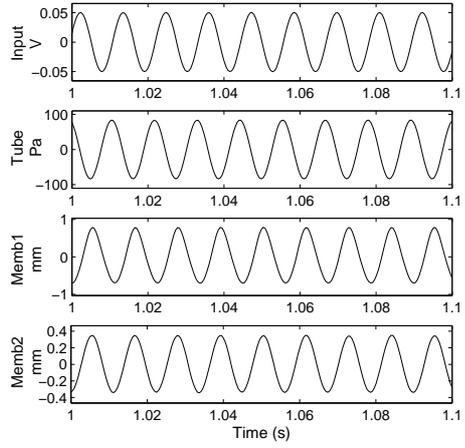}\label{for_1}}
\subfigure[${\cal A}=0.1\ V$.]{\includegraphics[width=6.5cm]{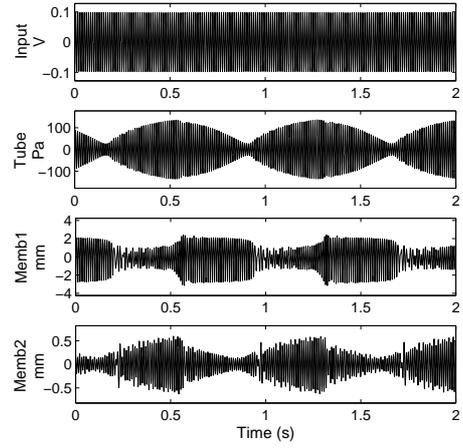}\label{for_2}}
\subfigure[${\cal A}=0.5\ V$.]{\includegraphics[width=6.5cm]{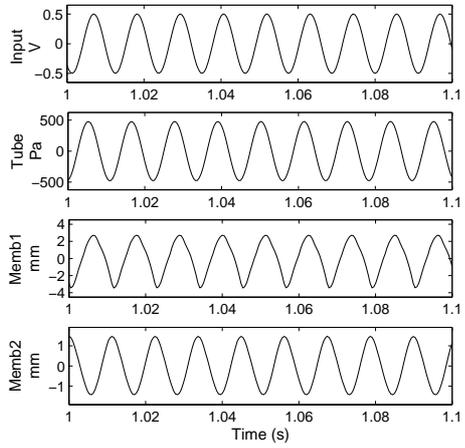}\label{for_3}}
\subfigure[${\cal A}=1\ V$.]{\includegraphics[width=6.5cm]{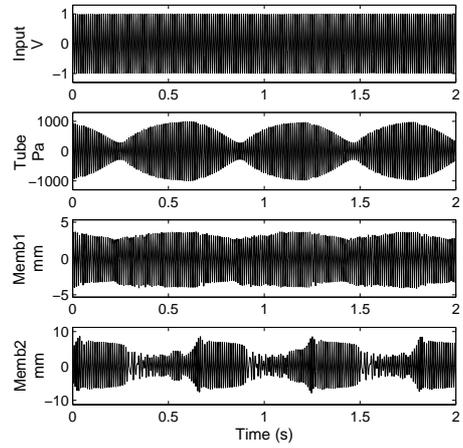}\label{for_4}}
\subfigure[${\cal A}=1.4\ V$.]{\includegraphics[width=6.5cm]{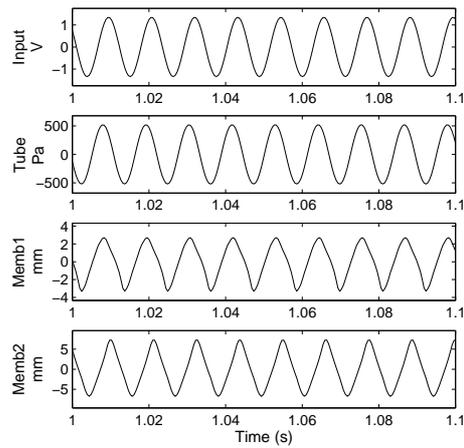}\label{for_5}}
\caption{\textit{Experimental results.} Oscillations of the system under sinusoidal excitation with different amplitudes and constant frequency (89 Hz). From top to bottom: sinusoidal input signal, acoustic pressure at the middle of the tube, displacement of the center of the membrane $M_1$ and displacement of the center of the membrane $M_2$.}
\label{for}
\end{figure}

The acoustic medium, composed by an open tube, has its first resonance at 89 Hz. As the aim of the membrane absorbers is to reduce this resonance, we describe the system behavior under sinusoidal excitation at the first resonance frequency of the acoustic medium. The figures from \ref{for_1} to \ref{for_5} show the time series observed experimentally with this excitation for a increasing input amplitude $\cal A$ using a small and thin membrane (membrane $M_1$ : $h_1=0.18$ mm, $R_1=2$ cm, $f_{11}=59$ Hz) and another larger and thicker (membrane $M_2$ : $h_2=0.39$ mm, $R_2=3$ cm, $f_{12}=47$ Hz). The system behavior can be described using the following steps, which define at the same time the first and second activation thresholds $S_1^{M_i}$ and $S_2^{M_i}$ of the membrane $M_i$:

\begin{enumerate}[label=\textbullet,itemsep=0pt]
\item $ {\cal A} < S_1^{M_1} $ (Fig. \ref{for_1}): sinusoidal regime where none of the two membranes is activated, both vibrating in opposite phase with the displacement at the end of the tube.
\item $ S_1^{M_1} < {\cal A} < S_2^{M_1} $ (Fig. \ref{for_2}): Activation of the first membrane leading to the quasi-periodic regime we already discussed in \cite{bellet10} involving the acoustic medium and the membrane $M_1$, the membrane $M_2$ remaining inactive and passively following the vibration inside the tube. During this quasi-periodic regime, each time the membrane $M_1$ is activated, a resonance capture occurs with the acoustic medium (the displacement of the membrane is in phase with the displacement at the end of the tube) and a targeted energy transfer is created from this linear primary system to the NES.
\item $ S_2^{M_1} < {\cal A} < S_1^{M_2} $ (Fig. \ref{for_3}): beyond the threshold $S_2^{M_1}$, the regime is periodic and the membrane $M_1$ vibrates with a large amplitude and in phase with the displacement at the end of the tube, while the membrane $M_2$ is still inactive and remains out of phase with this last degree of freedom. So far, behaviors are exactly the same as those observed with a single membrane set-up in \cite{bellet10}.
\item $ S_1^{M_2} < {\cal A} < S_2^{M_2} $ (Fig. \ref{for_4}): activation of the membrane $M_2$ leading to a second quasi-periodic regime involving the acoustic medium and the membrane $M_2$, while the membrane $M_1$ preserves its periodic evolution (barely disturbed by the quasiperiodic regime that exists around it).
\item $ {\cal A} > S_2^{M_2} $ (Fig. \ref{for_5}): beyond the last threshold $S_2^{M_2}$, the regime is periodic again and the membrane $M_2$ also vibrates with a large amplitude and in phase with the displacement at the end of the tube and the membrane $M_1$. Within this last regime, both membranes are on a resonance capture with the acoustic medium.
\end{enumerate}

Basically, these steps correspond to the successive activation of the different membranes, each one corresponding to the same phenomena as those observed with a single membrane set-up. Note that these mecanisms, where $S_2^{M_1} < S_1^{M_2}$, with clearly separated phases is a consequence of two very different membrane configurations (radius, thickness and pre-stress). For practical experimental limitations, the case of very close membrane configurations couldn't be studied on this set-up. This case will be discussed further in this paper thanks to simulations.

\subsection {Free oscillations}

\begin{figure}
\centering
\subfigure[${\cal A}=0.12\ V$.]{\includegraphics[width=13cm]{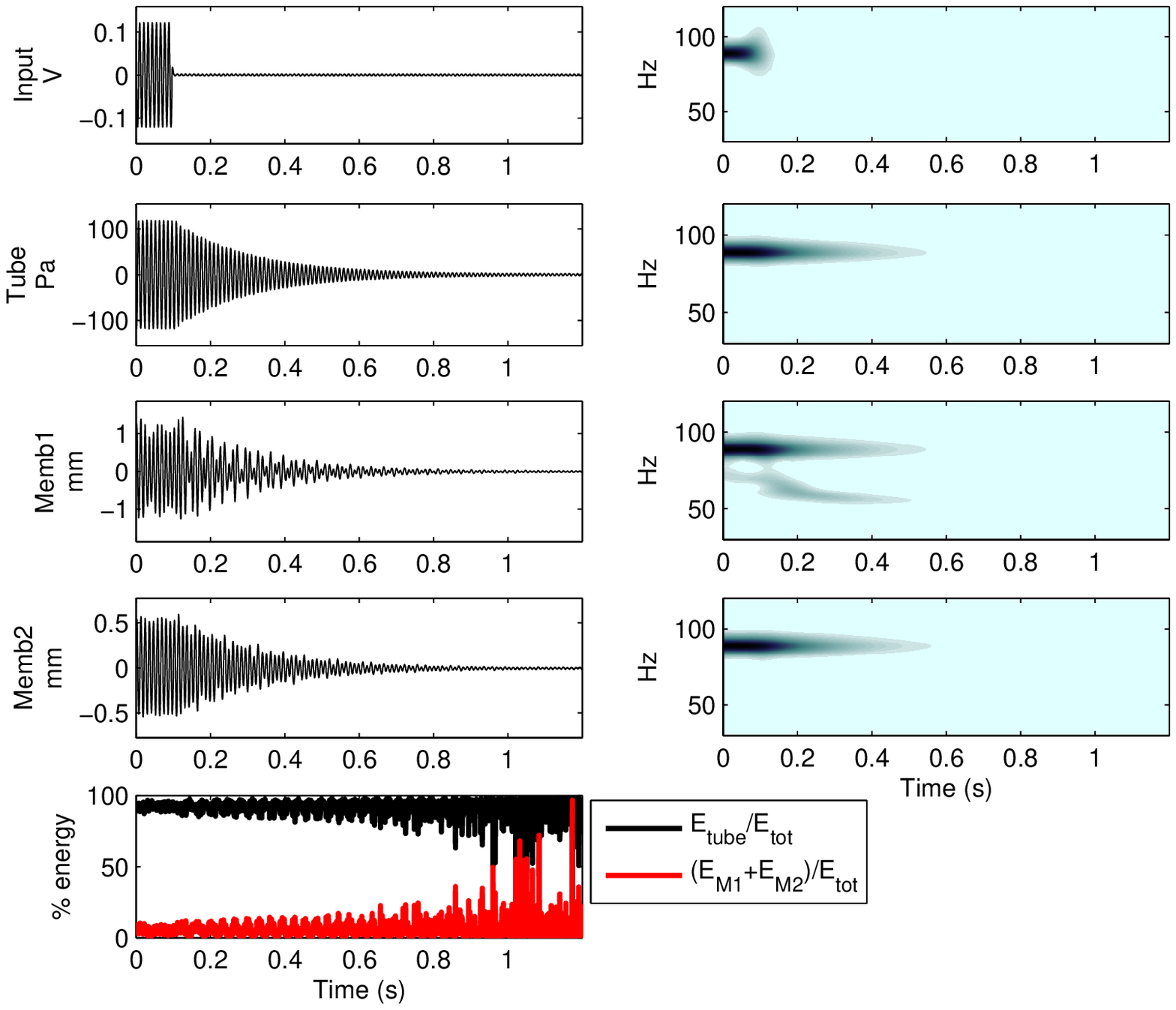}\label{lib_1}}
\subfigure[${\cal A}=0.4\ V$.]{\includegraphics[width=13cm]{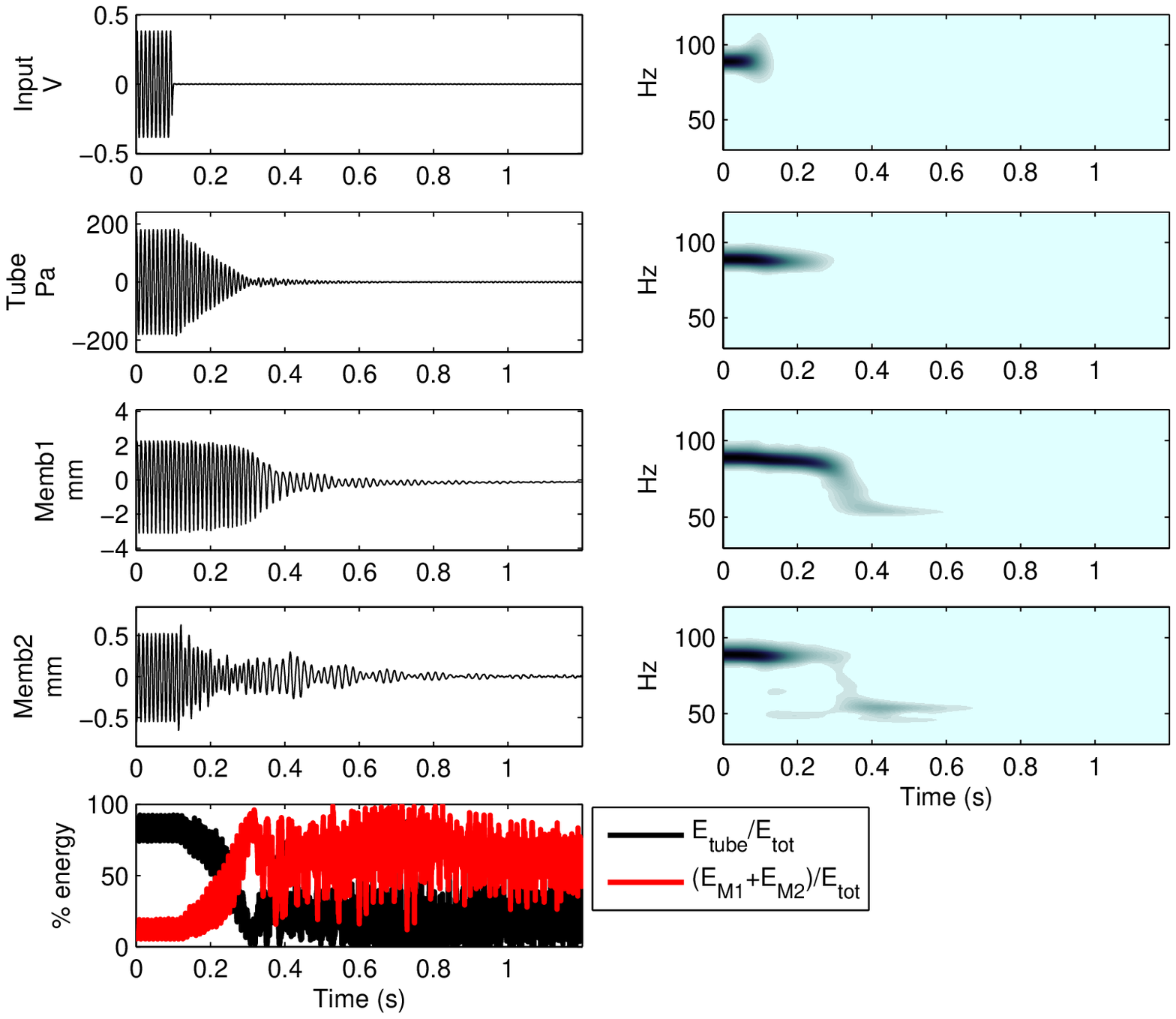}\label{lib_2}}
\end{figure}

\begin{figure}
\centering
\subfigure[${\cal A}=2\ V$.]{\includegraphics[width=13cm]{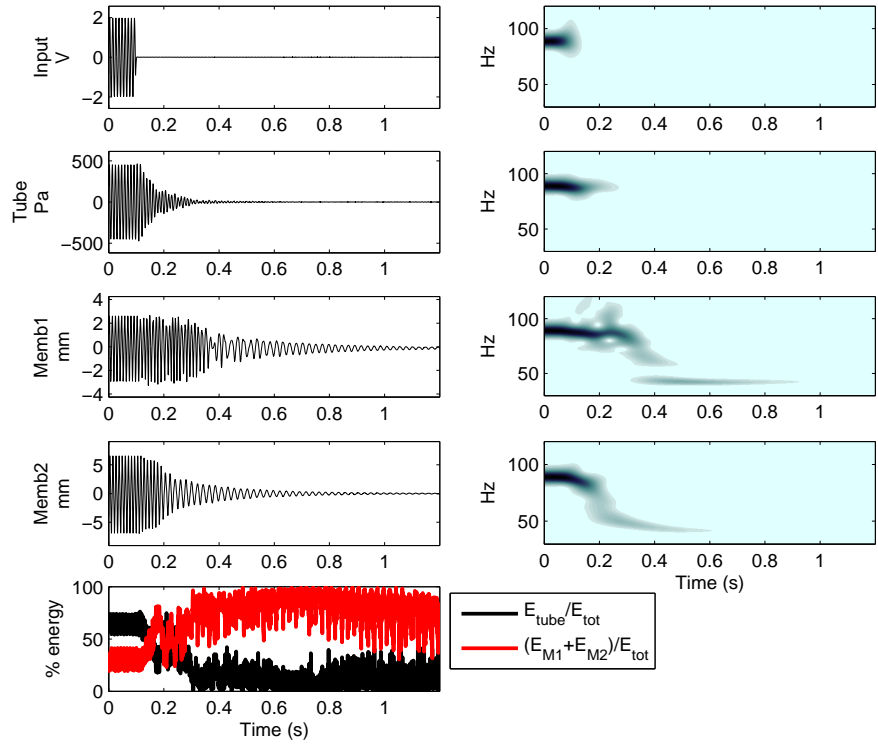}\label{lib_3}}
\caption{\textit{Experimental results.} Free oscillations of the system after different initial conditions. From top to bottom: input signal, acoustic pressure at the middle of the tube, displacement of the center of the membrane $M_1$, displacement of the center of the membrane $M_2$ and percentage of energy in the tube and in the membranes. On the right: wavelet transforms of the corresponding times series of the left side.}
\label{lib}
\end{figure}

\begin{figure}
\centering
\includegraphics[width=10cm]{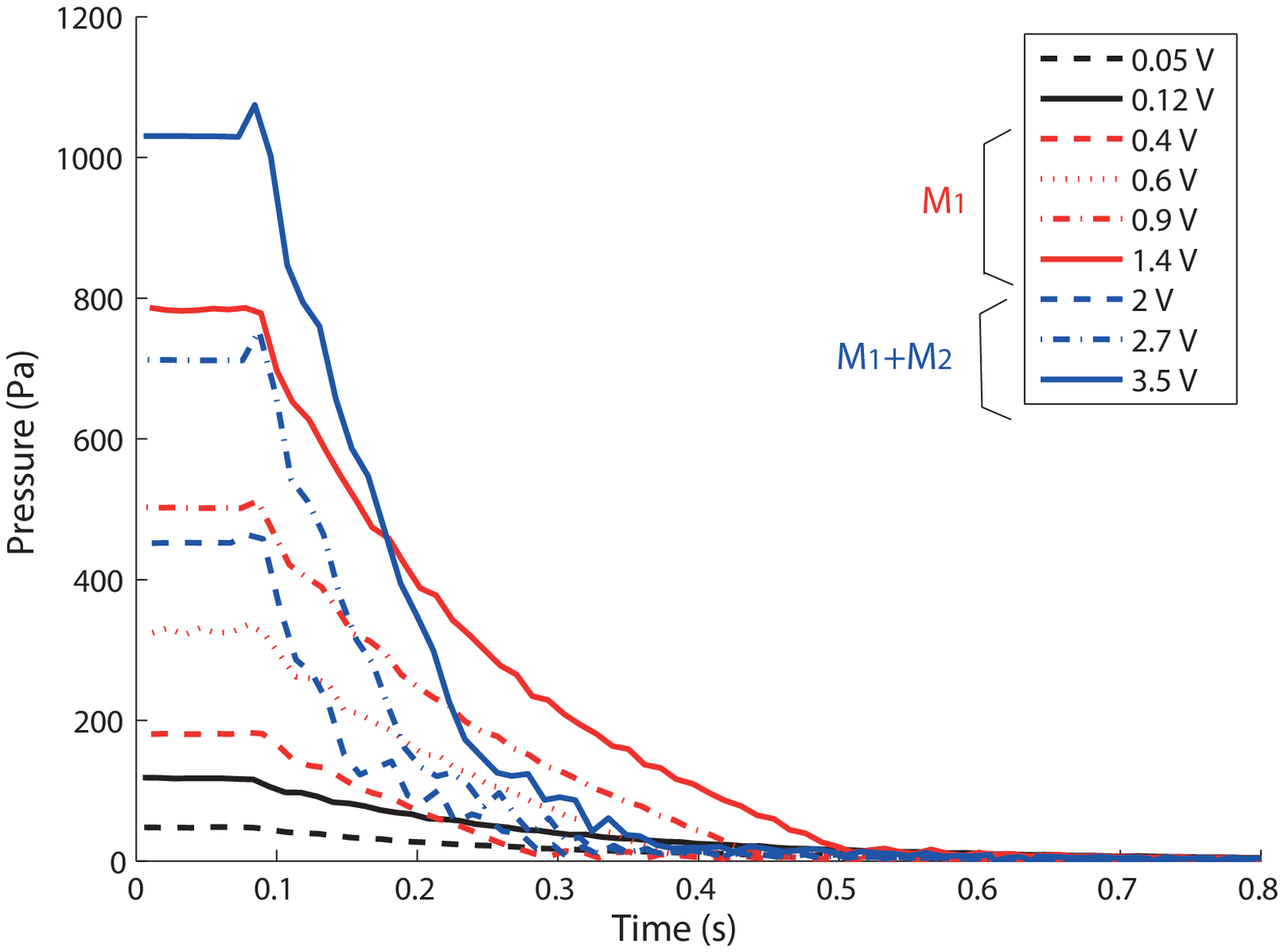}
\caption{\textit{Experimental results.} Positive envelopes of the sound pressure time series for all initial conditions.}
\label{env}
\end{figure}

The figures from \ref{lib_1} to \ref{lib_3} show the different types of free oscillations observed with these two membranes, after a sinusoidal excitation at 89 Hz suddenly stopped, for increasing values of initial input level. These figures include a diagram showing the evolution of the percentage of energy present in the acoustic medium and in both membranes. Wavelet transforms are also presented, in order to observe the temporal evolution of the frequency of the different time series. A description of the observed phenomena can be summarized as follows:

\begin{enumerate}[label=\textbullet,itemsep=0pt]
\item $ {\cal A} < S_1^{M_1} $ (Fig. \ref{lib_1}) : for low initial energy conditions, the membranes are not initially active and the decrease of the acoustic pressure is classically exponential. The energy remains always localized in the acoustic medium and the spectrum of the membranes movements mainly reflects a single component at the acoustic resonance frequency, the membranes just following passively the acoustic vibration.
\item $ S_2^{M_1} < {\cal A} < S_1^{M_2} $ (Fig. \ref{lib_2}) : beyond the second activation threshold of the membrane $M_1$, and below the first threshold of the membrane $M_2$, a targeted energy transfer occurs from the acoustic medium to the membrane $M_1$ and we observe an almost linear decrease similar to what we observed previously with a single membrane set-up. Indeed, since the membrane $M_2$ is not activated, it has no influence on the rest of the system and observations correspond to what has been seen so far in \cite{bellet10}. The energy diagram clearly shows how fast the energy is localized on the membrane, from the initial instant of free oscillations. After the transfer, the energy remains always localized on the membrane: the transfer is irreversible and the energy is then damped by viscosity in the membrane. On the time-frequency diagrams, we can see the resonance capture that the membrane $M_1$ only leaves when the acoustic level reaches zero, which avoids the return of the energy. After that, due to the energy-frequency dependence of a cubic oscillator, the frequency of the membrane decreases with its amplitude, until it reaches its first resonance frequency (59 Hz). After the cancellation of the sound, the frequency of the membrane $M_2$ follows passively the same variation than the frequency of membrane $M_1$, as it is then the last vibrating element in the system.
\item $ {\cal A} > S_2^{M_2} $ (Fig. \ref{lib_3}) : beyond the second threshold of the second membrane $M_2$, both membranes are initially activated on the resonance capture with the acoustic medium. The decrease is then still linear, but much faster. The presence of the second membrane allows a stronger targeted energy transfer if the initial energy is high enough. Note that below a certain acoustic level, during the decrease, the membrane $M_2$ stops acting, leaving the membrane $M_1$ proceeding the end of the transfer. The slope becomes then the same than before, corresponding to the previous case where a single membrane is active.
\end{enumerate}
The Fig. \ref{env} summarizes these steps showing the positive envelopes of the sound pressure time series for all initial conditions. It provides an alternative illustration of the difference between the three cases described above.

The successive activation of the membranes thus does not change much about the overall mechanism of targeted energy transfer, just making it more and more intense each time an additional membrane is activated. This additive mechanism is similar to what was obtained analytically in reference \cite{nguyen10}  (equation 4.51) for the case of a nongrounded configuration with several nonlinear absorbers in parallel.

\subsection {Frequency responses}

\begin{figure}
\centering
\includegraphics[width=10cm]{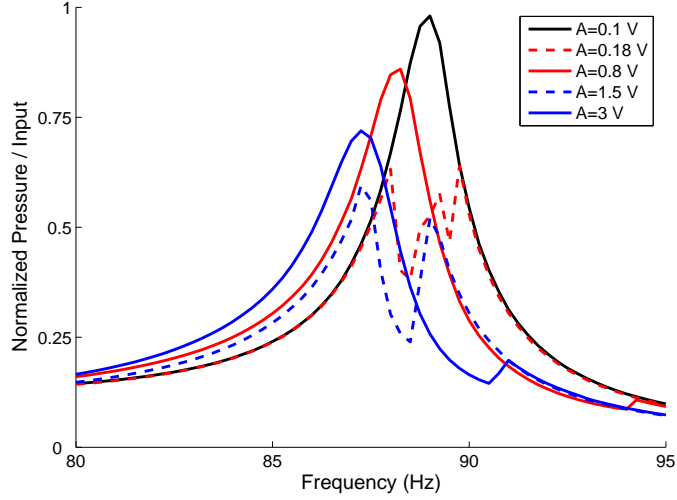}
\caption{\textit{Experimental results.} Frequency responses of the acoustic environment coupled to two membranes for different input amplitudes.}
\label{frf}
\end{figure}

This section finally deals with the influence of both membranes on the frequency responses of the acoustic medium. These responses have been measured with a swept sine source at constant amplitude. Five of them, corresponding to the five types of responses observed depending on the input level, are plotted in Fig. \ref{frf}:
\begin{enumerate}[label=\textbullet,itemsep=0pt]
\item ${\cal A}=0.1$ V: for low input levels, none of the two membranes is never activated and the response simply follows the first resonance peak of the acoustic medium.
\item ${\cal A}=0.18$ V: for levels between $S_1^{M_1}$ and $S_2^{M_1}$, i.e. high enough to observe the quasi-periodic regime of the membrane $M_1$, the resonance peak of the acoustic medium is clipped. This clipping is only due to the action of the first membrane, the level being too low for the membrane $M_2$ to act.
\item ${\cal A}=0.8$ V: for levels between $S_2^{M_1}$ and $S_1^{M_2}$, the frequency response has a new resonance peak slightly shifted to lower frequencies and whose maximal amplitude is slightly lower than the original peak. Along this peak, the membrane $M_1$ is activated, vibrates in phase with the acoustic displacement of the air at the end of the tube and acts as an additional mass on the system, whereas the membrane $M_2$ is still inactive, the level being still too weak for that membrane. These first three types of frequency responses are the same than what we have seen in the previous paper \cite{bellet10}, since for now only one membrane is active.
\item ${\cal A}=1.5$ V: for levels between $S_1^{M_2}$ and $S_2^{M_2}$, the membrane $M_2$ is activated and its quasiperiodic regime allows to clip this last second resonance peak. The membrane $M_2$ acts independently, on the resonance peak of a global system (acoustic medium and membrane $M_1$).
\item ${\cal A}=3$ V: for very high input levels, above $S_2^{M_2}$, a third resonance peak is observed, still more shifted to lower frequencies and whose maximal amplitude a little lower again. Along this peak, both membranes are activated and vibrate in phase with the acoustic displacement of the air at the end of the tube.
\end{enumerate}

In summary, we are still attending the successive action of the membranes that are activated one after the other, when the input level increases. They act independently of each other, each time repeating the same process, having an influence only on the resonance peak that they ``see'', creating a succession of clipping and new shifted resonance peaks, one membrane clipping the peak created by the previous one. There is no negative correlation effects, and this addition of NES can be seen as a way to enlarge the useful range of a TET system.

\section{Extension for more membranes thanks to numerical simulations}

The earlier model, which corresponds to a set-up with two membranes, can be extended to a $n$ membranes set-up. The resulting system is then:
\begin{eqnarray}\label{eq:systdim4m}
&&m_a \ddot{u}_a + c_f \dot{u}_a + k_a u_a + S_t k_b (S_t u_a - \sum_{j=1}^n \frac{S_j}{2} q_j) = F cos(\Omega t) \nonumber\\ 
&&m_1 \ddot{q}_1 + k_{11} [ (\frac{f_{11}}{f_{01}})^2 q_1 + \eta \dot{q}_1 ] + k_{31} [q^3_1 + 2 \eta q^2_1 \dot{q}_1 ] + \frac{S_1}{2} k_{b} (\sum_{j=1}^n \frac{S_j}{2} q_j - S_t u_a) = 0 \nonumber\\
&& ... \nonumber\\
&&m_i \ddot{q}_i + k_{1i} [ (\frac{f_{1i}}{f_{0i}})^2 q_i + \eta \dot{q}_i ] + k_{3i} [q^3_i + 2 \eta q^2_i \dot{q}_i ] + \frac{S_i}{2} k_{b} (\sum_{j=1}^n \frac{S_j}{2} q_j - S_t u_a) = 0 \nonumber\\
&& ... \nonumber\\
&&m_n \ddot{q}_n + k_{1n} [ (\frac{f_{1n}}{f_{0n}})^2 q_n + \eta \dot{q}_n ] + k_{3n} [q^3_n + 2 \eta q^2_n \dot{q}_n ] + \frac{S_n}{2} k_{b} (\sum_{j=1}^n \frac{S_j}{2} q_j - S_t u_a) = 0 \nonumber\\
&&\ \ \ \ \ \ \ \ \ \mbox{with}\ \ \ m_a=\frac{\rho_a S_t L}{2} \ ,\ \ m_i= \frac{\rho_m h S_i }{3} \ ,\ \ k_b=\frac{\rho_a c_0^2}{V_2} \ ,\ \ k_a=\frac{\rho_a S_t c_0^2 \pi^2}{2L} \nonumber\\
&&\ \ \ \ \ \ \ \ \ \ \ \ \ \ \ k_{1i}=\frac{1.015^4 \pi^5}{36}\frac{Eh_i^3}{(1-\nu^2)R_i^2} \ ,\ \ k_{3i}=\frac{8\pi E h_i}{3(1-\nu^2) R_i^2} \nonumber\\
&&\ \ \ \ \ \ \ \ \ \ \ \ \ \ \ f_{0i}=\frac{1}{2\pi} \sqrt{\frac{1.015^4 \pi^4}{12}\frac{Eh_i^2}{(1-\nu^2)\rho_m R_i^4}} \ ,\ \ S_i=\pi R_i^2 \nonumber
\end{eqnarray}

The simulations with two membranes showed the same phenomena as those observed experimentally, with an agreement similar to what we observed so far with a single membrane set-up in the paper \cite{bellet10}. In order to extend and generalize the previous results to a $n$ membranes set-up, we used this model and performed numerical simulations of the frequency responses for two configurations with four membranes, one with clearly different membranes and another with quasi-identical membranes.

\subsection{The case of four clearly different membranes}

The configurations of the membranes were chosen so that they have significantly different thresholds and in a well known order (see Table \ref{tab_config_diff}).
\begin{table}[h]
  \centering
  \begin{tabular}{|c|c|c|c|c|}
    \hline
    Membrane & $R_i$ (mm) & $h_i$ (mm) & $f_{1i}$ (Hz) & $\eta_i$ (s)\\
    \hline
    $M_1$ & 20 & 0.2 & 50 & 0.00025 \\
    $M_2$ & 30 & 0.3 & 40 & 0.00025 \\
    $M_3$ & 40 & 0.4 & 30 & 0.00025 \\
    $M_4$ & 50 & 0.5 & 20 & 0.00025 \\
    \hline
  \end{tabular}
  \caption{Configurations of the four different membranes. $R_i$, $h_i$, $f_{1i}$, $\eta_i$ are respectively the radius, thickness, first resonance frequency and damping factor of the membrane $M_i$.}
 \label{tab_config_diff}
\end{table}
\begin{figure}
\hspace{-1cm}
\includegraphics[width=16cm]{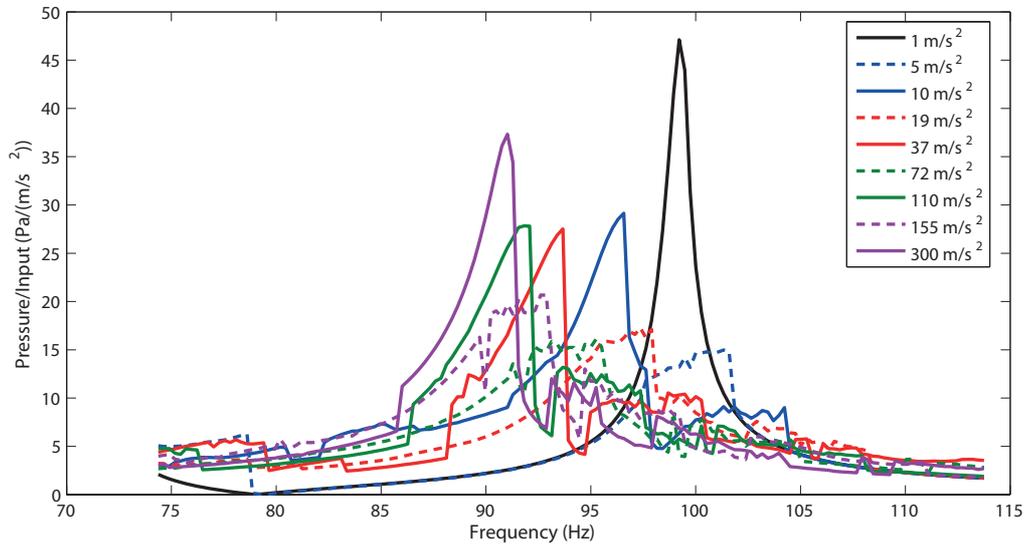}
\caption{\textit{Numerical results.} Frequency responses with four different membranes for nine different input levels.}
\label{frf4m}
\end{figure}
\begin{figure}
\centering
\includegraphics[width=10cm]{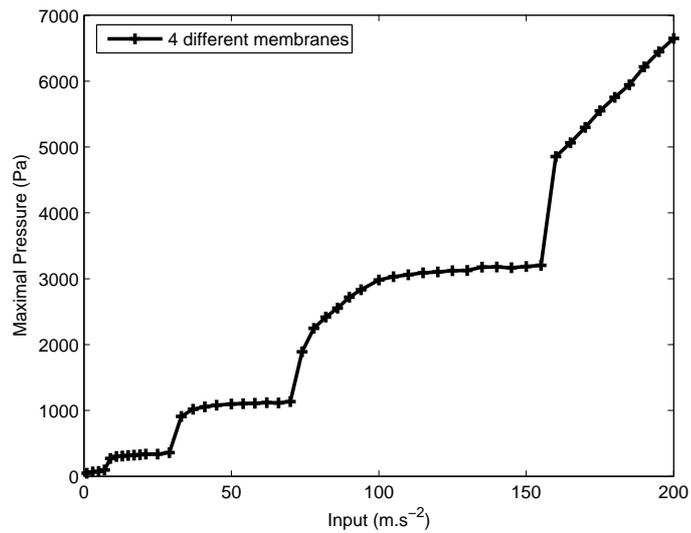}
\caption{\textit{Numerical results.} Curve connecting the maxima of each frequency response for a configuration with four different membranes in parallel.}
\label{crete4diff}
\end{figure}

The results of these simulations are shown in Fig. \ref{frf4m} where it appears that the mechanism observed experimentally remains similar and is simply extended to the case of four membranes:

\begin{enumerate}[label=\textbullet,itemsep=0pt]
\item For a very low input level, the frequency response follows the original peak of the acoustic medium.
\item From a certain input level, this peak is clipped by the first membrane while all others remain inactive.
\item Increasing the input level creates a new resonance peak slightly shifted to lower frequencies.
\item The same process is observed whenever a new membrane is activated: when the input level allows to activate an additional membrane, this enables a clipping of the last created peak, then a new peak, each time shifted to lower frequencies, appears increasing again the input level.
\end{enumerate}

So we observe in Fig. \ref{frf4m} nine different frequency responses, corresponding to nine different input levels: the original peak, the original peak clipped by $M_1$, the peak of the system \{tube+$M_1$\}, the peak of the system \{tube+$M_1$\} clipped by $M_2$, the peak of the system \{tube+$M_1$+$M_2$\}, the peak of the system \{tube+$M_1$+$M_2$\} clipped by $M_3$, the peak of the system \{tube+$M_1$+$M_2$+$M_3$\}, the peak of the system \{tube+$M_1$+$M_2$+$M_3$\} clipped by $M_4$ and finally the peak of the system \{tube+$M_1$+$M_2$+$M_3$+$M_4$\}.

The Fig. \ref{crete4diff} presents the curve connecting the maxima of each frequency response, each point showing the maximal pressure level observed for a certain input level. This curve highlights different horizontal zones, where a certain membrane acts as a nonlinear energy sink, preventing the acoustic level to exceed the level where this membrane gets activated. As the four membrane configurations are very different, these zones are here clearly separated and independent.

\subsection{The case of several very similar membranes}

In order to study the interest of using several very similar membrane, we performed different simulations of the behavior of the system with one, two, three and four membranes in parallel, whose configurations were chosen as mentioned in Table \ref{tab_config_same}.
\begin{table}[h]
  \centering
  \begin{tabular}{|c|c|c|c|c|}
    \hline
    Membrane & $R_i$ (mm) & $h_i$ (mm) & $f_{1i}$ (Hz) & $\eta_i$ (s)\\
    \hline
    $M_1$ & 30 & 0.3 & 40 & 0.00025 \\
    $M_2$ & 31 & 0.31 & 40 & 0.00025 \\
    $M_3$ & 32 & 0.32 & 40 & 0.00025 \\
    $M_4$ & 33 & 0.33 & 40 & 0.00025 \\
    \hline
  \end{tabular}
  \caption{Configurations of the four very similar membranes. $R_i$, $h_i$, $f_{1i}$, $\eta_i$ are respectively the radius, thickness, first resonance frequency and damping factor of the membrane $M_i$.}
 \label{tab_config_same}
\end{table}

\begin{figure}
\centering
\includegraphics[width=10cm]{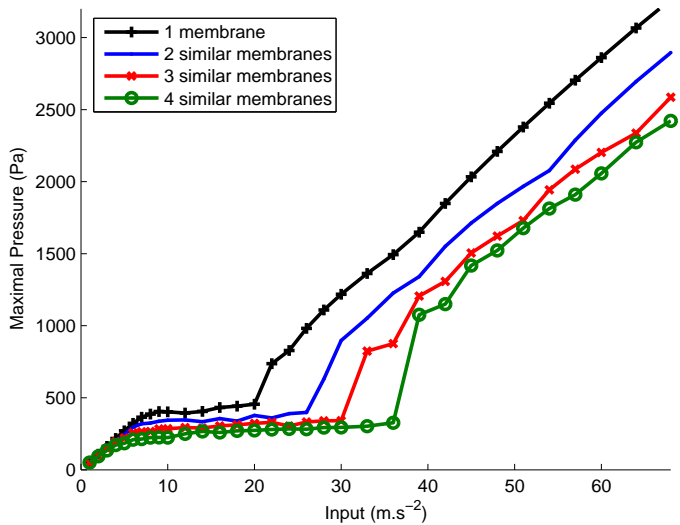}
\caption{\textit{Numerical results.} Curves connecting the maxima of each frequency response for four different configuration: a configuration with only one membrane, and three other configurations with one, two and three additional and very similar membranes in parallel.}
\label{crete4sim}
\end{figure}

The results of this study are shown in the Fig. \ref{crete4sim} where the curves of maximal acoustic level depending on the input level are plotted for these four different configurations. This figure shows that the horizontal zone where the targeted energy transfer exists can be extended thanks to an additional similar membrane. In each situation where we simulated several membranes, we could have replaced them with a single one. But this last membrane should have had properties unavailable in practice. As one of the most important drawbacks of nonlinear absorbers is to act only in a restricted zone, that is to say between two energy thresholds, the possibility of extending this zone is a very important result.

\section{Conclusion}

Based on the set-up of the previous paper \cite{bellet10}, this mainly experimental study of targeted energy transfer in acoustics thanks to two nonlinear membranes absorbers in parallel showed, in both temporal and frequency aspects, that the use of several NES is a way to extend the energy range and the efficiency of TET.
This is done thanks to an additional effect of the different membranes, which are activated in turn and behave relatively independently, repeating several times the same phenomena as those we know for a single nonlinear absorber set-up. Thanks to a model and numerical simulations, we could extend and generalize this results to a higher number of parallel absorbers: an addition of different membranes creates a new zone of TET and an additiona of similar ones extends the existing zone of TET. Since previous works about multiple NES only deal with a series coupling, because of specific constraints that are not existing in our acoustic case (mechanical systems in which the absorber is in a nongrounded configuration), this paper brings new information in this growing field of nonlinear absorbers, where acoustics has different properties than mechanical engineering applications of TET.

\section*{Aknowledgement}

This research was supported by French National Research Agency in the context of the ADYNO project (ANR-07-BLAN-0193).

\bibliographystyle{elsarticle-num}
\bibliography{biblio_papier_2mem}

\end{document}